\documentclass[12pt]{article}

\usepackage{amsmath,amssymb}
\usepackage{amsfonts}
\usepackage{graphicx}
\usepackage{a4wide}

\usepackage[T2A]{fontenc}
\usepackage[cp1251]{inputenc}
\usepackage[russian]{babel}

\begin{document}

\noindent {\sl Problems of Information Transmission},\\
\noindent vol. 56, no. 2, pp. 64-81, 2020.

\vskip 0.8cm

\begin{center}
{\large\bf New Upper Bounds in the Hypothesis Testing Problem with Information Constraints \footnote[1]
{The reported study was funded by RFBR according to the research
project 19-01-00364.}}
\end{center}

\begin{center} {\bf M. V. Burnashev}
\end{center}

\begin{center}\emph{Kharkevich Institute for Information Transmission Problems, \\
Russian Academy of Sciences, Moscow, Russia \\
email}: burn@iitp.ru
\end{center}

{\begin{quotation} \small 
We consider a hypothesis testing problem where a part of data cannot
be observed. Our helper observes the missed data and can send us
a limited amount of information about them. What kind of this
limited information will allow us to make the best statistical
inference? In particular, what is the minimum information sufficient
to obtain the same results as if we directly observed all the data?
We derive estimates for this minimum information and some other
similar results.
\end{quotation}}

{\begin{quotation} \small
\end{quotation}}

\begin{center}
{\large\bf \S\;1. Introduction and main results}
\end{center}

{\bf 1. Statement of the problem}. Similarly to \cite{BAH1, BAH2},
a binary symmetric channel {\rm BSC}$(p)$ on length $n$, with
unknown crossover probability $p$ is considered. In order to
distinguish input and output alphabets $E^{n} = \{0,1\}^{n}$,
denote them $E_{\rm in}^{n}$ and $E_{\rm out}^{n}$, respectively.
Concerning the value $p$, there are two hypotheses (one of them is
true) : $H_{0}: p = p_{0}$ and $H_{1}: p = p_{1}$, where
$0 < p_{0}, p_{1} \leq 1/2$.

Denote by ${\mathbf P}$ and ${\mathbf Q}$ conditional output
distributions on the {\rm BSC}$(p)$ output for hypotheses $H_{0}$
and $H_{1}$, respectively. Then probabilities to get the output
block $\boldsymbol{y}=(y_{1},\ldots,y_{n})$ provided the input block
$\boldsymbol{x}=(x_{1},\ldots,x_{n})$ are given by
$$
\begin{gathered}
{\mathbf P}(\boldsymbol{y}|\boldsymbol{x})=
(1-p_{0})^{n-d(\boldsymbol{x},\boldsymbol{y})}
p_{0}^{d(\boldsymbol{x},\boldsymbol{y})}, \qquad
{\mathbf Q}(\boldsymbol{y}|\boldsymbol{x})=
(1-p_{1})^{n-d(\boldsymbol{x},\boldsymbol{y})}
p_{1}^{d(\boldsymbol{x},\boldsymbol{y})},
\end{gathered}
$$
where $d(\boldsymbol{x},\boldsymbol{y})$ - the Hamming distance
between blocks $\boldsymbol{x}$ and $\boldsymbol{y}$ (i.e. the number
of non-coincident components of those vectors on length $n$).

The following problem of minimax testing of hypotheses $H_{0}$ and
$H_{1}$ is considered. We (i.e. ``the statistician'') observe only
the channel output block $\boldsymbol{y} \in E_{\rm out}^{n}$,
while our ``helper'' observes only the channel input
$\boldsymbol{x} \in E_{\rm in}^{n}$. It is assumed that we do not
have any prior information on the input block $\boldsymbol{x}$.
Clearly, that based only on the output block $\boldsymbol{y}$ we
are not able to make any reasonable conclusions on unknown value $p$.

Assume further that for a prescribed value $R > 0$, our helper is
allowed to partition in advance the input space
$E_{\rm in}^{n} = \{0,1\}^{n}$ on $N \leq 2^{Rn}$ arbitrary parts
$\{X_{1}, \ldots, X_{N}\}$, and to inform us (in some additional way)
to which part $X_{i}$ belongs the input block $\boldsymbol{x}$.
Clearly, only the case $N < 2^{n}$, i.e. $R < 1$, is interesting
(otherwise, the helper can simply inform us on the block
$\boldsymbol{x}$).

For example, the helper may transmit to the statistician exact
values of the first $Rn$ components $x_{1}, \ldots, x_{Rn}$
(but inform nothing on the next values $x_{i}$). Such simple
partitioning of the input space $E_{\rm in}^{n}$ (on cylinder sets
$\{X_{i}\}$), generally speaking, is not optimal. From the
statistician point of view input data $(x_{1}, \ldots, x_{n})$
represent very strong nuisance parameter.

We may also say that the optimal limited information on the block
$\boldsymbol{x}$ means the optimal ``contraction'' of full
information on the block $\boldsymbol{x}$. Of course, such optimal
``contraction'' depends on prior information on transfer probability
$p$ and a quality criteria used.

{\it Remark}\, 1. Clearly, the problem will not be changed if
the statistician observes the channel input, and the helper
observes the channel output.

Based on observation $\boldsymbol{y}$ and the index $i$ of the part
$X_i$ the statistician makes
a decision in favor of one of hypotheses $H_0$ or $H_1$. In order
to avoid overcomplification we consider only nonrandomized decision
methods (then the problem essence and results remain the same).

We consider partitions $\{X_{1},\ldots,X_{N}\}$ and decision
methods that are asymptotically (as $n \to \infty$) optimal.
Similar, but much more general problem statements were considered,
for example, in \cite{AC1,HK1,AB,HA1,HSA1,WAT17}.

{\it Remark}\, 2. As far as we know, all results in that area
(see, for example, \cite{BAH1, BAH2,AC1,HK1,AB,HA1,HSA1,WAT17})
have the form: ``it is possible to get the following testing
performance \ldots ''. Our aim is to get an opposite result, i.e.
to show that ``it is impossible to get a better
result than \ldots''.

Below we denote $\log x = \log_2 x$. For a finite set $A$ we denote
by $|A|$ its cardinality. Introduce balls and spheres in $E^{n}$
\begin{equation} \label{defB}
\begin{gathered}
{\bf B}_{\boldsymbol{x}}(p) = \{\boldsymbol{u}:
d(\boldsymbol{x},\boldsymbol{u}) \leq pn\}, \qquad
\boldsymbol{x},\boldsymbol{u} \in E^{n}, \\
{\bf S}_{\boldsymbol{x}}(p) = \{\boldsymbol{u}:
d(\boldsymbol{x},\boldsymbol{u}) = pn\}.
\end{gathered}
\end{equation}

{\bf 2. Error probability exponents and dual problem}. Let a
partition $\{X_{1}, \ldots, X_{N}\}$ of the input space
$E_{\rm in}^{n} = \{0,1\}^{n}$ be chosen. Then general decision
making can be described as follows. For each partition element
$X_{i}$ we choose a set ${\mathcal A}(X_i) \subset E_{\rm out}^{n}$,
and based on observation $\boldsymbol{y}$ and known element $X_{i}$,
make a decision
(${\mathcal A}^{c} = E_{\rm out}^{n} \setminus {\mathcal A}$):
$$
\begin{gathered}
\boldsymbol{y} \in {\mathcal A}(X_i) \Longrightarrow H_{0}; \qquad
\boldsymbol{y} \in {\mathcal A}^{c}(X_i) \Longrightarrow H_{1}.
\end{gathered}
$$

Assume that we set a partition $\{X_{1}, \ldots, X_{N}\}$ of
the input space $E_{\rm in}^{n} = \{0,1\}^{n}$. For each partition
element $X_i$ we choose a set
${\mathcal A}(X_i) \subset E_{\rm out}^{n}$, and based on
observation $\boldsymbol{y}$ and known element $X_i$ make a decision
(${\mathcal A}^{c} = E_{\rm out}^{n} \setminus {\mathcal A}$):
$$
\boldsymbol{y} \in {\mathcal A}(X_i) \Longrightarrow H_0; \quad
\boldsymbol{y} \in {\mathcal A}^{c}(X_i) \Longrightarrow H_1.
$$
Define error probabilities of the 1--kind $\alpha_{n}$
and the 2--kind $\beta_{n}$ as
$$
\begin{gathered}
\alpha_n = {\rm Pr}(H_1|H_0) = \max_{i=1,\ldots,N}
\max_{\boldsymbol{x} \in X_i}
{\mathbf P}\left({\mathcal A}^{c}(X_i)|\boldsymbol{x}\right), \\
\beta_n = {\rm Pr}(H_0|H_1) = \max_{i=1,\ldots,N}
\max_{\boldsymbol{x} \in X_i}
{\mathbf Q}\left({\mathcal A}(X_i)|\boldsymbol{x}\right).
\end{gathered}
$$

Let $\gamma \geq 0$ - a given constant. We demand that the 1--kind
error probability $\alpha_{n}$ satisfies the condition
\begin{equation} \label{def1}
\alpha_{n} = {\rm Pr}(H_{1}|H_{0}) \leq 2^{-\gamma n}.
\end{equation}

We are interested in the minimal possible (over all partitions
$\{X_{i}\}$ of the input space $E_{\rm in}^{n}$
and all decisions) 2--kind error probability $\inf\beta_{n}$.
We investigate the asymptotic case as $n \to \infty$ and $N = 2^{Rn}$, where
$0 < R < 1$ -- a given constant.
\footnote{In order to simplify formulas we don't use integer part
sign of value $2^{Rn}$} Then for the best partition $\{X_{i}\}$
and decision methods denote
\begin{equation} \label{def2}
e(\gamma, R) = \lim_{n \to \infty}
\frac{1}{n} \log_2 \frac{1}{\inf\beta_{n}} > 0,
\end{equation}
where $\inf$ is taken over all partitions $\{X_{i}\}$ and decision
methods satisfying the condition \eqref{def1}.

Our main aim is upperbounds for the function $e(\gamma, R)$
(see lowerbounds in \cite{BAH1}). In the paper we limit ourselves
to the case $\gamma \to 0$, evaluating the function $e(0,R)=e(R)$,
and the related function $r_{\rm crit}(p_{0},p_{1})$ (that case
sometimes is called Neiman-Pierson problem). In other paper we
will consider the case $\gamma > 0$.

It will be convenient for us to consider also the equivalent dual
problem (without the helper). Let a value $r$, $0 < r < 1$
be given, and we may choose any set
${\mathcal X} \subset E_{\rm in}^{n}$ of $X = 2^{rn}$ input blocks.
It is known also that the input block $\boldsymbol{x}$ belongs to the
chosen set $\mathcal X$. We observe the channel output
$\boldsymbol{y}$ and, knowing the set $\mathcal X$, consider the
testing of hypotheses $H_0$ and $H_1$ problem. We choose a set
${\mathcal A}$ and depending on observation $\boldsymbol{y}$ make
the decision:
$$
\begin{gathered}
\boldsymbol{y} \in {\mathcal A} \Longrightarrow H_0; \quad
\boldsymbol{y} \in {\mathcal A}^{c} \Longrightarrow H_1.
\end{gathered}
$$
Define 1--kind $\alpha_{n}$ and 2--kind $\beta_{n}$ error
probabilities as
$$
\alpha_n = \max_{\boldsymbol{x} \in {\mathcal X}}
{\mathbf P}\left({\mathcal A}^c|\boldsymbol{x}\right), \quad
\beta_n = \max_{\boldsymbol{x}\in {\mathcal X}}
{\mathbf Q}\left({\mathcal A}|\boldsymbol{x}\right).
$$

Assume that for the 1-kind error probability $\alpha_{n}$
condition \eqref{def1} is fulfilled, and we want to choose the set
${\mathcal X}\subset E_{\rm in}^{n}$ of cardinality $X = 2^{rn}$
and decision method in order to achieve the minimal possible
2-kind error probability $\inf\beta_{n}$.
Similarly to \eqref{def2},
for such dual problem define the function $e_{\rm d}(\gamma, r)$
\begin{equation} \label{def2a}
e_{\rm d}(\gamma, r) = \lim_{n \to \infty}
\frac{1}{n} \log_2 \frac{1}{\min\beta_{n}} > 0,
\end{equation}
where minimum  is taken over all sets
${\mathcal X}\subset E_{\rm in}^{n}$ of cardinality $X = 2^{rn}$
and all decision methods.

The following result establishes simple relation between functions
$e(\gamma, R)$ and $e_{\rm d}(\gamma, r)$.

{P r o p o s i t i o n \ 1} \cite[Proposition 1]{BAH1}.
{\it The following relation holds true}
\begin{equation} \label{rel1}
e(\gamma, 1 - R)= e_{\rm d}(\gamma, R); \qquad
0 \leq R \leq 1, \quad \gamma \geq 0.
\end{equation}

By virtue of Proposition 1 and the formula \eqref{rel1} it is
sufficient to investigate the function $e_{\rm d}(\gamma, r)$.
In the paper we limit ourselves to the case $\gamma \to 0$,
investigating  the function $e_{\rm d}(0,r)$.

{\it Remark } 3. Essentially, we consider the case when
distributions $P(x,y)$ and $Q(x,y)$ have the
form: $P(x,y)=p(x)P(y|x)$ and $Q(x,y)=p(x)Q(y|x)$.

{\bf 3. Known input block}. Assume that we know the input block
$\boldsymbol{x}$ (then we may set $\boldsymbol{x} = \boldsymbol{0}$)
and we observe the output block $\boldsymbol{y}$. If we demand only
$\alpha_{n} \to 0$, $n \to \infty$ (i.e. $\gamma=0$), and we are
interested only in the exponent (on $n$) of 2-kind error
probability $\beta_{n}$, then as $n \to \infty$ by Central Limit
Theorem and Pearson-Neiman lemma the optimal decision set in favor
of $H_{0}$ (i.е. $p_{0}$) is the spherical slice
${\mathbf B}_{\boldsymbol{0}}(p_{0}+\delta) \setminus
{\mathbf B}_{\boldsymbol{0}}(p_{0}-\delta)$ in $E_{\rm out}^{n}$
(see \eqref{defB}), where $\delta > 0$ - small. Then for the exponent
(on $n$) of 2-kind error probability $\beta_{n}$ we have
$$
\begin{gathered}
\frac{1}{n}\log\beta_{n} = \frac{1}{n}\log\left[
\binom{n}{p_{0}n}(1-p_{1})^{(1-p_{0})n}p_{1}^{p_{0}n}\right] + o(1),
\qquad n \to \infty,
\end{gathered}
$$
and therefore we get as $n \to \infty$
\begin{equation} \label{known1}
\begin{gathered}
\frac{1}{n}\log\frac{1}{\beta_{n}} = -(1-p_{0})\log(1-p_{1}) -
p_{0}\log p_{1}-h(p_{0}) + o(1)= D(p_{0}||p_{1}) + o(1),
\end{gathered}
\end{equation}
where
\begin{equation} \label{defD}
\begin{gathered}
D(a||b) =a\log\frac{a}{b} +(1-a)\log\frac{1-a}{1-b}.
\end{gathered}
\end{equation}

{\it Remark} 4. The function $D(a||b)$ is the divergence for two
binomial random variables with parameters $a$ and $b$, respectively.
In other words, it gives the best possible exponent for 2--kind
error probability provided fixed 1--kind error probability (i.e. its
exponent equals $0$), when testing two simple hypotheses:
$H_{0}:p=a$ versus $H_{1}: p = b$.

With $\gamma =r= 0$ for the value $e_{\rm d}(\gamma,0)$
(see \eqref{def2a}) we have from \eqref{known1}
\begin{equation} \label{def2b}
e_{\rm d}(0,0) = D(p_{1}||p_{0}).
\end{equation}

{\bf 4. Unknown input block and critical rate}.
If we know the input block $\boldsymbol{x}$  and $\alpha_{n} \to 0$,
then the best exponent  $e_{\rm d}(0,0)$ for 2--kind error
probability $\beta_{n}$ is given by the formula \eqref{def2b}.

If we know only that the input block $\boldsymbol{x}$ belongs
to the set $\mathcal X$ of cardinality $X \sim 2^{rn}$, then for the
best such set $\mathcal X$ the exponent $e_{\rm d}(0, r)$
of 2-kind error probability $\beta_{n}$
is defined by the formula \eqref{def2a}. It is clear that
\begin{equation} \label{r1}
e_{\rm d}(\gamma, r) \leq e_{\rm d}(\gamma, 0),
\quad \gamma \geq 0, \quad 0 \leq r \leq 1.
\end{equation}

The function $e_{\rm d}(\gamma, r)$ does not increase in $r$.
Then the following natural question arises: does there exist
$r(\gamma)>0$ for which the equality in \eqref{r1} holds, and, if
so, what is the maximal rate $r_{\rm crit}(\gamma)$ ? Limiting
ourselves to the case $\gamma=0$, define the critical rate
$r_{\rm crit}(p_{0},p_{1}) = r_{\rm crit}(p_{0},p_{1},0)$ as
(see \eqref{def2b})
\begin{equation} \label{r2}
r_{\rm crit} = r_{\rm crit}(p_{0},p_{1})=
\sup\{r: e_{\rm d}(0,r)= e_{\rm d}(0,0) =D(p_{0}||p_{1})\}.
\end{equation}

In other words, what is the maximal cardinality $2^{rn}$ of the best
set $\mathcal X$ for which we can achieve the same asymptotic
efficiency as for known input block $\boldsymbol{x}$ (although we
don't know the input block $\boldsymbol{x}$) ?

Similarly, introduce the critical rate $R_{\rm crit}$ for
the original problem (see \eqref{def2})
\begin{equation} \label{r3}
R_{\rm crit}(p_{0},p_{1})= \inf\{R: e(0,R)=e(0,1)=D(p_{0}||p_{1})\}.
\end{equation}

By virtue of Proposition 1 and \eqref{r3} we have
\begin{equation} \label{r4}
R_{\rm crit}(p_{0},p_{1}) =1 - r_{\rm crit}(p_{0},p_{1}).
\end{equation}

The paper main result is

Т е о р е м а \,1. {\it If $p_{1} < p_{0} \leq 1/2$, then there
exists $p_{1}^{*}(p_{0}) \leq p_{0}$, such that for any
$p_{1} \leq p_{1}^{*}(p_{0})$ the formula holds}
\begin{equation} \label{Theorem1}
\begin{gathered}
r_{\rm crit}(p_{0},p_{1}) = 1- R_{\rm crit}(p_{0},p_{1}) =
1-h(p_{0}), \qquad 0 < p_{1} \leq p_{1}^{*}< p_{0} \leq 1/2.
\end{gathered}
\end{equation}

{\it Remark }5. Although the value $r_{\rm crit}(p_{0},p_{1})$ in
\eqref{Theorem1} coincides with the channel {\rm BSC}$(p_{0})$
capacity, its origin \eqref{r2} is related with the function
$e_{\rm d}(0,r)$, similar to the channel reliability function
$E(r,p)$ in information theory \cite{E, G2}. Exact form of
the reliability function $E(r,p)$ is only partially  known
\cite{Bur15}. For that reason, in the proof of Theorem 1 rather
recent results on spectrum of binary codes are used
(as in \cite{Bur15,Bur6,Bur5}). Complete description of the function
$e_{\rm d}(\gamma, r)$ looks rather difficult problem.

In \S 2 the lower bound for $r_{\rm crit}$ (Proposition 2) is
presented. In \S 3 the general formula for 2-kind error probability
$\beta_{n}$ (Lemma 1) is derived. Using the method of ``two
hypotheses'', in \S 4 Theorem 1 is proved. But generally speaking,
the upper bound \eqref{Theorem1} for $r_{\rm crit}$ is weaker than
the corresponding lower bound from \S 2. In \S 5 using additional
combinatoric arguments one more upper bound for $r_{\rm crit}$
(Proposition 3) is derived. In \S 6 the accuracy of the lower bound
for $r_{\rm crit}$ from Proposition 2 is shown, provided some
additional condition is fulfilled. In Appendix some necessary
analytic results are presented.

Below in the paper $f \sim g$ means
$n^{-1}\ln f = n^{-1}\ln g + o(1),\,n \to \infty$, and
$f \lesssim g$ means
$n^{-1}\ln f \leq n^{-1}\ln g + o(1),\,n \to \infty$.

\begin{center}
{\large\bf \S\;2. Lower bound for $r_{\rm crit}$}
\end{center}

Next result follows from \cite[Proposition 2]{BAH1}.

{P r o p o s i t i o n \,2}.
{\it For $r_{\rm crit}(p_{0},p_{1})$ lower bounds hold
\begin{equation} \label{prop21}
\begin{gathered}
r_{\rm crit}(p_{0},p_{1}) \geq 1-h(p_{0}), \qquad \text{if} \quad
0 < p_{1} < p_{0} \leq 1/2.
\end{gathered}
\end{equation}
and}
\begin{equation} \label{prop2}
\begin{gathered}
r_{\rm crit}(p_{0},p_{1}) \geq 1-h(p_{0}) - D(p_{0}||p_{1}),
\qquad \text{if} \quad 0 < p_{0} < p_{1} \leq 1/2.
\end{gathered}
\end{equation}

{\sl Proof}.
For given $r$, $0 < r < 1$, choose randomly and equiprobably
a set $\mathcal X$ of $X = 2^{rn}$ input blocks $\boldsymbol{x}$.
It was shown in \cite[Proposition 2]{BAH1} that if
$p_{0} < p_{1} \leq 1/2$, then for any $\tau$,
$p_0 \leq \tau \leq p_1$, there exist a set $\mathcal X$ and
a decision method for which the following inequalities hold
\begin{equation} \label{low1b}
\begin{gathered}
\frac{1}{n}\log\frac{1}{\alpha_{n}} \geq D(\tau||p_{0}), \qquad
\frac{1}{n}\log\frac{1}{\beta_{n}} \geq
\min\{D(\tau||p_{1}),1-h(\tau)-r\}.
\end{gathered}
\end{equation}
If it is sufficient to have $\alpha_{n} \to 0$, $n \to \infty$,
then setting in \eqref{low1b} $\tau = p_{0}$, we get \eqref{prop2}
from \eqref{r2}.

If $p_{1} < p_{0} \leq 1/2$, then changing $p_{0}$ with $p_{1}$
and $\alpha_{n}$ with $\beta_{n}$ in \eqref{low1b} then for any
$\tau$ we have
\begin{equation} \label{low1c}
\begin{gathered}
\frac{1}{n}\log\frac{1}{\alpha_{n}} \geq
\min\{D(\tau||p_{0}),1-h(\tau)-r\}, \qquad
\frac{1}{n}\log\frac{1}{\beta_{n}} \geq D(\tau||p_{1}).
\end{gathered}
\end{equation}
If it is sufficient to have $\alpha_{n} \to 0$, $n \to \infty$,
then setting $\tau = p_{0}$ in \eqref{low1c}, from
\eqref{r2} we get \eqref{prop21}. $\qquad \Box$

\begin{center}
{\large\bf \S\;3. General formula for 2-kind error probability
$\beta_{n}$}.
\end{center}

Let
${\cal C}_{n}(r)= \{\boldsymbol{x}_{1},\ldots,\boldsymbol{x}_{M}\}$ -
a set (code) of $M=2^{rn}$ different input codeblocks. For the code
${\cal C}_{n}(r)$ and 1-kind error probability $\alpha_{n}$ denote
by ${\cal D}_{0}=
{\cal D}_{0}({\cal C}_{n},\alpha_{n}) \subseteq E_{\rm out}^{n}$
the optimal decision set in favor of $H_{0}$, minimizing 2-kind
error probability $\beta_{n}$. Although the set ${\cal D}_{0}$ has
rather complicated form, it is possible to establish some its
properties sufficient for proving Theorem 1.

Set a small $\delta>0$ and for each $\boldsymbol{x}_{k}$,
$k=1,\ldots,M$, introduce the spherical slide in $E_{\rm out}^{n}$
\begin{equation} \label{defSL}
\begin{gathered}
SL_{\boldsymbol{x}_{k}}(p_{0},\delta) =
{\mathbf B}_{\boldsymbol{x}_{k}}(p_{0}+\delta) \setminus
{\mathbf B}_{\boldsymbol{x}_{k}}(p_{0}-\delta) = \{\boldsymbol{u}:
|d(\boldsymbol{x}_{k},\boldsymbol{u})-p_{0}n| \leq \delta n\},
\end{gathered}
\end{equation}
where $B_{\boldsymbol{x}}(p)$ is defined in \eqref{defB}.
For each $\boldsymbol{x}_{k}$ introduce also the set
\begin{equation} \label{proberB7c}
\begin{gathered}
D_{\boldsymbol{x}_k}(\delta) = {\cal D}_{0}\bigcap
SL_{\boldsymbol{x}_k}(p_{0},\delta).
\end{gathered}
\end{equation}

Since we need $\alpha_{n} \to 0$, $n \to \infty$, the optimal set
${\cal D}_{0}$ contains an ``essential'' part of each set
$SL_{\boldsymbol{x}_{k}}(p_{0},\delta)$, $k=1,\ldots,M$. In order
to evaluate it, note that for any $\boldsymbol{x}_{k}$ and
$\boldsymbol{u},\boldsymbol{z} \in
SL_{\boldsymbol{x}_k}(p_{0},\delta)$ we have
\begin{equation} \label{cond1d}
\begin{gathered}
\frac{{\mathbf P}(\boldsymbol{u}|p_{0},\boldsymbol{x}_{k})}
{{\mathbf P}(\boldsymbol{z}|p_{0},\boldsymbol{x}_{k})} =
\left(\frac{q_{0}}
{p_{0}}\right)^{d(\boldsymbol{z},\boldsymbol{x}_{k})-
d(\boldsymbol{u},\boldsymbol{x}_{k})} \leq
\left(\frac{q_{0}}{p_{0}}\right)^{2\delta n}, \qquad q_{0}= 1-p_{0}.
\end{gathered}
\end{equation}
By Chebychev exponential inequality (Chernov bound) for any
$\boldsymbol{x}_{k}$ and small $\delta > 0$ we get
\begin{equation} \label{Theor2c}
\begin{gathered}
\log {\mathbf P}\{\boldsymbol{u}\not \in
SL_{\boldsymbol{x}_{k}}(p_{0},\delta)|\boldsymbol{x}_{k},p_{0}\}
\leq -\frac{n\delta^{2}}{2p_{0}q_{0}}.
\end{gathered}
\end{equation}
Then by \eqref{defSL}, \eqref{proberB7c} and \eqref{Theor2c} we
have for any $\boldsymbol{x}_{k}$
\begin{equation} \label{proberB7d}
\begin{gathered}
{\mathbf P}\left\{D_{\boldsymbol{x}_{k}}(\delta)|
p_{0},\boldsymbol{x}_{k}\right\} \geq 1-{\mathbf P}
\left\{\boldsymbol{u}\not\in {\cal D}_{0}|p_{0},\boldsymbol{x}_{k}
\right\} - {\mathbf P}\left\{\boldsymbol{u} \not\in
SL_{\boldsymbol{x}_{k}}(p_{0},\delta)
|p_{0},\boldsymbol{x}_{k}\right\} \geq  \\
\geq 1-\alpha_{n}- e^{-n^{2}\delta^{2}/(2p_{0}q_{0})},
\end{gathered}
\end{equation}
and by \eqref{cond1d} also have
\begin{equation} \label{cond1e}
\begin{gathered}
\delta_{1}|SL_{\boldsymbol{x}_k}(p_{0},\delta)| \leq
|D_{\boldsymbol{x}_k}(\delta)| \leq
|SL_{\boldsymbol{x}_k}(p_{0},\delta)|, \\
\delta_{1} =
\left(1-\beta_{n}-e^{-n^{2}\delta^{2}/(2p_{0}q_{0})}\right)
\left(\frac{p_{0}}{q_{0}}\right)^{2\delta n}.
\end{gathered}
\end{equation}
Since $D_{\boldsymbol{x}_k}(\delta) \subseteq {\cal D}_{0}$ for
any $\boldsymbol{x}_{k}$, then by \eqref{proberB7c},
\eqref{proberB7d} and \eqref{cond1e} for the probability
${\mathbf P}(e|p_{1},\boldsymbol{x}_{i})$ we have
\begin{equation} \label{cond1f}
\begin{gathered}
{\mathbf P}(e|p_{1},\boldsymbol{x}_{i}) = {\mathbf P}
\{{\cal D}_{0}|p_{1},\boldsymbol{x}_{i}\} \sim
{\mathbf P}\left\{\bigcup_{k=1}^{M}D_{\boldsymbol{x}_k}(\delta)|
p_{1},\boldsymbol{x}_{k}\right\} \sim  \\
\geq \delta_{1}{\mathbf P}\left\{\bigcup_{k=1}^{M}
SL_{\boldsymbol{x}_k}(p_{0},\delta)|
p_{1},\boldsymbol{x}_{i}\right\}.
\end{gathered}
\end{equation}

For $t > 0$ and each $\boldsymbol{x}_{i}$ introduce the set
\begin{equation} \label{defDx1}
\begin{gathered}
D_{\boldsymbol{x}_{i}}(t,p) = \left\{\boldsymbol{u}:
\begin{array}{ccc}
 \text{there exists } \boldsymbol{x}_{k} \neq \boldsymbol{x}_{i},
 \text{ such that} \\
  d(\boldsymbol{x}_{i},\boldsymbol{u}) = tn,
d(\boldsymbol{x}_{k},\boldsymbol{u})=pn
                  \end{array}
\right\}.
\end{gathered}
\end{equation}

{L e m m a\; 1}. {\it For $2$-kind error probability $\beta_{n}$ of
a code
${\cal C}_{n}= \{\boldsymbol{x}_{1},\ldots,\boldsymbol{x}_{M}\}$
and the optimal set ${\cal D}_{0}$ in favor of $H_{0}$, the formula
holds as $n \to \infty$ }
\begin{equation} \label{lemma1}
\begin{gathered}
\frac{\log\beta_{n}}{n} \sim \max_{t > 0}\left\{\frac{1}{n}
\log\left[\frac{1}{M}\sum_{i=1}^{M}
|D_{\boldsymbol{x}_{i}}(t,p_{0})|\right] +
t\log p_{1} + (1-t)\log (1-p_{1})\right\}.
\end{gathered}
\end{equation}
{\it The critical rate $r_{\rm crit}(p_{0},p_{1})$ is defined by
the formula} ($M = 2^{rn}$)
\begin{equation} \label{lemma1a}
\begin{gathered}
r_{\rm crit}(p_{0},p_{1}) =
\sup\left\{r: F(p_{0},p_{1},r) \leq 0\right\} =
\inf\left\{r: F(p_{0},p_{1},r) > 0\right\},
\end{gathered}
\end{equation}
{\it where}
\begin{equation} \label{lemma1aa}
\begin{gathered}
F(p_{0},p_{1},r) = \lim_{n \to \infty}\min_{|{\cal C}_{n}|\leq M}
\max_{t}F(p_{0},p_{1},r,{\cal C}_{n},t), \\
F(p_{0},p_{1},r,{\cal C}_{n},t) = \frac{1}{n}\log\left[
\sum_{i=1}^{M}|D_{\boldsymbol{x}_{i}}(t,p_{0})|\right] +
(p_{0}-t)\log\frac{1-p_{1}}{p_{1}} -r - h(p_{0}).
\end{gathered}
\end{equation}

{\sl Proof}. Using \eqref{cond1f} with $\delta = o(1)$ and
$\delta_{1} = e^{o(n)}$ as $n \to \infty$, we have
\begin{equation} \label{lemma1ad}
\begin{gathered}
\beta_{n} = \max_{i}{\mathbf P}(e|p_{1},\boldsymbol{x}_{i}) \sim
\frac{1}{M}\sum_{i=1}^{M}{\mathbf P}
(e|p_{1},\boldsymbol{x}_{i}) \sim \frac{\delta_{1}}{M}
\sum_{i=1}^{M}{\mathbf P}\left\{\bigcup_{k=1}^{M}
SL_{\boldsymbol{x}_k}(p_{0},\delta)\Big|
p_{1},\boldsymbol{x}_{i}\right\}.
\end{gathered}
\end{equation}
From \eqref{defDx1} and \eqref{lemma1} for each $\boldsymbol{x}_{i}$
\begin{equation} \label{Pe2ab}
\begin{gathered}
{\mathbf P}
\left\{\bigcup_{k =1}^{M}SL_{\boldsymbol{x}_k}(p_{0},\delta)\Big|
p_{1},\boldsymbol{x}_{i}\right\} \sim {\mathbf P}
\left\{\bigcup_{t > 0}D_{\boldsymbol{x}_{i}}(t,p_{0})\Big|
p_{1},\boldsymbol{x}_{i}\right\} \sim  \\
\sim \max_{t > 0}\left\{p_{1}^{tn}(1-p_{1})^{(1-t)n}
|D_{\boldsymbol{x}_{i}}(t,p_{0}|\right\}.
\end{gathered}
\end{equation}
Therefore from \eqref{lemma1ad} and \eqref{Pe2ab} the formula
\eqref{lemma1} follows.

Since
$$
\begin{gathered}
{\mathbf P}\left\{SL_{\boldsymbol{x}_i}(p_{0},\delta)|
p_{1},\boldsymbol{x}_{i}\right\} \sim
{\mathbf P}\left\{d(\boldsymbol{x}_{i},\boldsymbol{u}) \geq p_{0}n|
p_{1},\boldsymbol{x}_{i}\right\} \sim 2^{-D(p_{0}||p_{1})n},
\end{gathered}
$$
the right-hand side of \eqref{lemma1} increases with $r$
(i.e. with $M=2^{rn}$), starting from $(-D(p_{1}||p_{0}))$.
Therefore, from \eqref{known1} and \eqref{lemma1} it follows that
the critical rate $r_{\rm crit}$ is the maximal rate $r$, such that
\begin{equation} \label{Pe2bb}
\begin{gathered}
\min_{\{\boldsymbol{x}_{i}\}}\max_{t > 0}\left\{\frac{1}{n}
\log\left[\sum_{i=1}^{M}|D_{\boldsymbol{x}_{i}}(t,p_{0})|\right] +
t\log p_{1} + (1-t)\log (1-p_{1})\right\}-r \leq -D(p_{0}||p_{1}).
\end{gathered}
\end{equation}
Note that
\begin{equation} \label{Pe2bd}
\begin{gathered}
D(p_{0}||p_{1}) + t\log p_{1} + (1-t)\log (1-p_{1}) =
-h(p_{0}) + (p_{0}-t)\log\frac{1-p_{1}}{p_{1}}.
\end{gathered}
\end{equation}
From \eqref{Pe2bb} and \eqref{Pe2bd} the formulas
\eqref{lemma1a}-\eqref{lemma1aa} follow. $\qquad \Box$

In particular, from \eqref{ineq1c} with $t = p_{0}$ we have
$$
\begin{gathered}
F(p_{0},p_{1},r,{\cal C}_{n},p_{0}) = o(1), \qquad n \to \infty.
\end{gathered}
$$

The main difficulty in analysis of relations
\eqref{lemma1a}-\eqref{lemma1aa} constitutes estimation of
cardinalities $|D_{\boldsymbol{x}_{i}}(t,p_{0})|$ in
\eqref{lemma1aa}, which depend on the code ${\cal C}_{n}$ geometry.
Similar problem arose in \cite{Bur15,Bur6,Bur5}, where the reliability
function $E(R,p)$ of the channel {\rm BSC}$(p)$ was investigated.
Direct estimation of those cardinalities leads to quite bulky
formulas.

\begin{center}
{\large\bf \S\;4. Upper bound for $r_{\rm crit}$: two hypotheses}.
\end{center}

We get a simple (but not very accurate) upper bound for
$r_{\rm crit}(p_{0},p_{1})$, using quite popular in mathematical
statistics (mainly, in estimation theory) method of ``two
hypotheses''. Using the formula \eqref{lemma1}, choose from the code
${\cal C}_{n}(r) = \{\boldsymbol{x}_{1},\ldots,\boldsymbol{x}_{M}\}$,
$M=2^{rn}$, any two codewords, say,
$\boldsymbol{x}_{1}$ и $\boldsymbol{x}_{2}$ with
$d(\boldsymbol{x}_{1},\boldsymbol{x}_{2}) = \omega n$. We may assume
that for a rate $r > 0$ the value $\omega$ satisfies constraints
$$
\begin{gathered}
0 < \omega \leq \omega_{\rm min}(r),
\end{gathered}
$$
where the value $\omega_{\rm min}(r)$ will be defined later. Replace
the code ${\cal C}_{n}(r)$ by the code ${\cal C}'$ of two chosen
codewords ${\cal C}'=\{\boldsymbol{x}_{1},\boldsymbol{x}_{2}\}$.
Then $\beta_{n}({\cal C}) \geq \beta_{n}({\cal C}')$. Similarly to
\eqref{lemma1ad}-\eqref{Pe2ab} we have
$$
\begin{gathered}
\beta_{n}({\cal C}') \sim 2^{-D(p_{0}||p_{1})n} +
{\mathbf P}\left\{SL_{\boldsymbol{x}_2}(p_{0},\delta)\Big|
p_{1},\boldsymbol{x}_{1}\right\}.
\end{gathered}
$$
We are interested when for $\boldsymbol{x}_{1},\boldsymbol{x}_{2}$
the following inequality holds
\begin{equation} \label{dist1b}
\begin{gathered}
\frac{1}{n}\log{\mathbf P}\left\{SL_{\boldsymbol{x}_2}(p_{0},\delta)
\Big|p_{1},\boldsymbol{x}_{1}\right\} > - D(p_{0}||p_{1}).
\end{gathered}
\end{equation}
Evaluate the probability in the left-hand side of \eqref{dist1b}.
For $d(\boldsymbol{x}_{i},\boldsymbol{x}_{k}) =\omega n$ denote
\begin{equation} \label{defD4}
\begin{gathered}
S_{\boldsymbol{x}_{i},\boldsymbol{x}_{k}}(t,p,\omega) =
\{\boldsymbol{u}: d(\boldsymbol{x}_{i},\boldsymbol{u}) = tn,
d(\boldsymbol{x}_{k},\boldsymbol{u})=pn,
d(\boldsymbol{x}_{i},\boldsymbol{x}_{k}) =\omega n\}.
\end{gathered}
\end{equation}
Then (see Appendix)
\begin{equation} \label{defD3a}
\begin{gathered}
\frac{1}{n}
\log|S_{\boldsymbol{x}_{i},\boldsymbol{x}_{k}}(t,p,\omega)|
= g(t,p,\omega) + o(1), \qquad n \to \infty, \\
\frac{1}{n}\log{\mathbf P}\left\{
S_{\boldsymbol{x}_{i},\boldsymbol{x}_{k}}(t,p,\omega)
\Big|p_{1},\boldsymbol{x}_{i}\right\} = g(t,p,\omega) -
t\log\frac{1-p_{1}}{p_{1}} + \log(1-p_{1}) + o(1),
\end{gathered}
\end{equation}
where $g(t,p,\omega)$ is defined in $\eqref{defg}$.
Therefore as $n \to \infty$ (see \eqref{geom2a1}-\eqref{geom2a})
\begin{equation} \label{geom2aa}
\begin{gathered}
\frac{1}{n}\log{\mathbf P}
\left\{SL_{\boldsymbol{x}_2}(p_{0},\delta)\Big|
p_{1},\boldsymbol{x}_{1}\right\} =
\frac{1}{n}\max_{t}\log{\mathbf P}\left\{
S_{\boldsymbol{x}_{1},\boldsymbol{x}_{2}}(t,p_{0},\omega)
\Big|p_{1},\boldsymbol{x}_{1}\right\} + o(1) = \\
= f(p_{0},p_{1},\omega)+ o(1)
\end{gathered}
\end{equation}
where
\begin{equation} \label{geom2aa1}
\begin{gathered}
f(p_{0},p_{1},\omega) = \max_{t}f(p_{0},p_{1},\omega,t), \\
f(p_{0},p_{1},\omega,t) = g(t,p_{0},\omega) -
t\log\frac{1-p_{1}}{p_{1}} + \log(1-p_{1}).
\end{gathered}
\end{equation}
We have
\begin{equation} \label{geom2aab}
\begin{gathered}
f'_{t}(p_{0},p_{1},\omega,t) = \log\frac{\omega-t}{t} -
\log\frac{p_{0}+t-\omega}{1-p_{0}-t}-2\frac{1-p_{1}}{p_{1}}, \qquad
f''_{tt}(p_{0},p_{1},\omega,t) < 0.
\end{gathered}
\end{equation}
By \eqref{Pe2bd} and \eqref{defD3a}-\eqref{geom2aa1} the inequality
\eqref{dist1b} takes the form
\begin{equation} \label{dist1g}
\begin{gathered}
\max_{t}F(p_{0},p_{1},\omega,t) > 0,
\end{gathered}
\end{equation}
where
\begin{equation} \label{dist1g1}
\begin{gathered}
F(p_{0},p_{1},\omega,t)= f(p_{0},p_{1},\omega,t)+ D(p_{0}||p_{1}) =
g(t,p_{0},\omega) +(p_{0}-t)\log\frac{1-p_{1}}{p_{1}} -h(p_{0}).
\end{gathered}
\end{equation}

If for some $p_{0},p_{1}$ and $\omega$ the inequality \eqref{dist1g}
holds, then the appropriate upper bound
\eqref{prop21}-\eqref{prop2} is valid. Denote by
$t_{1}^{0} = t_{1}^{0}(p_{0},p_{1},\omega)$ the maximizing value
$t$ in \eqref{geom2aa1} (it remains the maximizing one in
\eqref{dist1g} as well). Then
\begin{equation} \label{geom2ac}
\begin{gathered}
f(p_{0},p_{1},\omega) =
f(p_{0},p_{1},\omega,t_{1}^{0}(p_{0},p_{1},\omega)).
\end{gathered}
\end{equation}
From the equation $f'_{t}(p_{0},p_{1},\omega,t) =0$ for $t_{1}^{0}$
from \eqref{geom2aab} we get
\begin{equation} \label{geom2ab}
\begin{gathered}
t_{1}^{0} = t_{1}^{0}(p_{0},p_{1},\omega) = \frac{{\sqrt{1+
(v_{0}-1)[(\omega-p_{0})^{2}v_{0}-(1-\omega-p_{0})^{2}+1]}-1}}
{v_{0}-1}, \\
\qquad v_{0}(p_{1}) = \left(\frac{1-p_{1}}{p_{1}}\right)^{2} \geq 1.
\end{gathered}
\end{equation}
Then from \eqref{dist1g1} and \eqref{geom2ab} we have
\begin{equation} \label{dist1e}
\begin{gathered}
F(p_{0},p_{1},\omega,t_{1}^{0}) = g(t_{1}^{0},p_{0},\omega) +
(p_{0}-t_{1}^{0})\log\frac{1-p_{1}}{p_{1}} -h(p_{0}).
\end{gathered}
\end{equation}

It is possible to check that for the function
$F(p_{0},p_{1},\omega,t_{1}^{0})$ from \eqref{dist1e} we have \\
$F(p_{0},p_{1},0,t_{1}^{0}) = 0$ and
$F''_{\omega\omega} < 0$, $\omega> 0$. Therefore, it is sufficient
to check the inequality \eqref{dist1g} with $t=t_{1}^{0}$ only for
the minimal value $\omega$  for the code ${\cal C}_{n}(r)$ (i.e.
for its code distance $d({\cal C})$).

Let $\omega_{\rm min}(r)n$ - the maximal possible code distance of
${\cal C}_{n}(r)$. For the value $\omega_{\rm min}(r)$ the following
bound is known \cite[formula (1.5)]{M4}
\begin{equation} \label{geom2ah}
\begin{gathered}
r \leq h\left[\frac{1}{2} -
\sqrt{\omega_{\rm min}(1-\omega_{\rm min})}\right], \qquad
\omega_{\rm min}=\omega_{\rm min}(r).
\end{gathered}
\end{equation}

Consider two possible cases 1) $p_{1} < p_{0} \leq 1/2$ and
2) $p_{0} < p_{1} \leq 1/2$.

{\bf 1. Case $p_{1} < p_{0} \leq 1/2$}. Setting
$r = 1-h(p_{0})$, denote by $\omega_{0} =\omega_{0}(p_{0})$ the
root of the equation (see \eqref{geom2ah})
$$
\begin{gathered}
1-h(p_{0}) = h\left[\frac{1}{2} - \sqrt{\omega(1-\omega)}\right].
\end{gathered}
$$
Then the inequality \eqref{dist1g} takes the form
($\omega_{0} =\omega_{0}(p_{0})$)
\begin{equation} \label{dist1h}
\begin{gathered}
F(p_{0},p_{1},\omega_{0},t_{1}^{0}) =
g(t_{1}^{0},p_{0},\omega_{0}) +
(p_{0}-t_{1}^{0})\log\frac{1-p_{1}}{p_{1}} -h(p_{0}) > 0.
\end{gathered}
\end{equation}

It is possible to check (Maple), that the inequality \eqref{dist1h}
is satisfied, if $p_{1} \leq p_{1}^{*}(p_{0})$, where
$$
\begin{gathered}
\begin{array}{ccccccccc}
  p_{0} & 0.1 & 0.12 & 0.15 & 0.2 & 0.3 & 0.4 & 0.45 & 0.49 \\
  p_{1}^{*}(p_{0}) & 0.0003 & 0.003 & 0.016 & 0.056 &
  0.17 & 0.317 & 0.4 & 0.48
\end{array}
\end{gathered}
$$

If $p_{0} \leq 0.20707$ (i.e.  $\omega < 0.273$), then in
\cite[формула (1.4)]{M4} there is a little bit more accurate than
\eqref{geom2ah} bound (but much more bulky).

{\bf 2. Case $p_{0} < p_{1} \leq 1/2$}.
It is possible to check (Maple), that the inequality \eqref{dist1g}
is not satisfied for any $p_{0} < p_{1}$ !

\begin{center}
{\large\bf \S\;5. Upper bound for $r_{\rm crit}$: combinatorics}.
\end{center}

We will get one more upper bound for $r_{\rm crit}$, based on the
same formula \eqref{lemma1}, but using additional  combinatorics
arguments.

{\bf 1. Combinatorics lemma}. In the code
${\cal C}_{n}= \{\boldsymbol{x}_{i}\}$ we call
$(\boldsymbol{x}_{i},\boldsymbol{x}_{j})$ $\omega$-pair, if
$d(\boldsymbol{x}_{i},\boldsymbol{x}_{j})= \omega n$. The total
number of $\omega$--pairs in a code ${\cal C}_{n}$ equals
$MB_{\omega n}$ (see \eqref{defBi}). We say that a point
$\boldsymbol{y} \in E^{n}$ is $(\omega,p,t)$--covered, if there
exists $\omega$--pair $(\boldsymbol{x}_{i},\boldsymbol{x}_{j})$
such that $d(\boldsymbol{x}_{i},\boldsymbol{y}) = pn$,
$d(\boldsymbol{x}_{j},\boldsymbol{y})= tn$. Denote by
$K(\boldsymbol{y},\omega,p,t)$ the number of
$(\omega,p,t)$--coverings of the point $\boldsymbol{y}$
(taking into account multiplicity of coverings), i.e.
\begin{equation}\label{defK}
K(\boldsymbol{y},\omega,p,t) =|\left\{(\boldsymbol{x}_{i},
\boldsymbol{x}_{j}): d(\boldsymbol{x}_{i},\boldsymbol{x}_{j}) =
\omega n,\, d(\boldsymbol{x}_{i},\boldsymbol{y}) = pn, \
d(\boldsymbol{x}_{j},\boldsymbol{y})= tn\right\}|,
\quad \omega > 0.
\end{equation}
Introduce sets (see \eqref{defDx1})
\begin{equation} \label{defD1}
\begin{gathered}
D_{\boldsymbol{x}_{i}}(t,p,\omega) =
\bigcup_{\boldsymbol{x}_{k}}
S_{\boldsymbol{x}_{i},\boldsymbol{x}_{k}}(t,p,\omega) = \\
= \left\{\boldsymbol{u}: \begin{array}{ccc}
 \text{there exists } \boldsymbol{x}_{k} \text{ such that }
d(\boldsymbol{x}_{i},\boldsymbol{x}_{k}) = \omega n,  \\
  d(\boldsymbol{x}_{i},\boldsymbol{u}) = tn,
d(\boldsymbol{x}_{k},\boldsymbol{u})=pn
                  \end{array}
\right\}.
\end{gathered}
\end{equation}
Then
$$
\begin{gathered}
D_{\boldsymbol{x}_{i}}(t,p) = \bigcup_{\omega > 0}
D_{\boldsymbol{x}_{i}}(t,p,\omega).
\end{gathered}
$$

For $t > 0$ introduce the value
\begin{equation} \label{defmy}
\begin{gathered}
m_{t}(\boldsymbol{y}) = \{\text{число }
\boldsymbol{x}_{i} \in {\bf S}_{\boldsymbol{y}}(t)\}.
\end{gathered}
\end{equation}
Then for any $\boldsymbol{y},p,t > 0$
\begin{equation} \label{defKmy}
\begin{gathered}
K(\boldsymbol{y},t,p)=m_{t}(\boldsymbol{y})m_{p}(\boldsymbol{y}).
\end{gathered}
\end{equation}

{L e m m a \; 2}. {\it For a code $\{\boldsymbol{x}_{i}\}$ and
$\omega,p,t > 0$ the formula holds} ({\it see \eqref{defK} и }
\eqref{defD1})
\begin{equation}\label{ineq1a}
\begin{gathered}
\sum_{i=1}^{M}|D_{\boldsymbol{x}_{i}}(t,p,\omega)| \leq
\sum_{\boldsymbol{y}\in E^{n}}K(\boldsymbol{y},\omega,t,p).
\end{gathered}
\end{equation}
{\it Also, if} ({\it see} \eqref{defmy})
\begin{equation}\label{ineq1cond1}
\begin{gathered}
\max_{\boldsymbol{y}}m_{p}(\boldsymbol{y}) = 2^{o(n)}, \qquad
n \to \infty,
\end{gathered}
\end{equation}
{\it then for any $\omega, t > 0$ }
\begin{equation}\label{ineq1b}
\begin{gathered}
\sum_{i=1}^{M}|D_{\boldsymbol{x}_{i}}(t,p,\omega)| =
2^{o(n)}\sum_{\boldsymbol{y}\in E^{n}}K(\boldsymbol{y},\omega,t,p),
\qquad n \to \infty.
\end{gathered}
\end{equation}

{\sl Proof}. Let $\boldsymbol{y}\in E^{n}$ and there are $m$ ordered
pairs $(\boldsymbol{x}_{i},\boldsymbol{x}_{j})$ with
$d(\boldsymbol{x}_{i},\boldsymbol{x}_{j})=\omega n$ and
$d(\boldsymbol{x}_{i},\boldsymbol{y}) = tn$,
$d(\boldsymbol{x}_{j},\boldsymbol{y})= pn$. Those $m$ pairs
$(\boldsymbol{x}_{i},\boldsymbol{x}_{j})$ have $m_{1} \leq m$
different first arguments $\{\boldsymbol{x}_{i}\}$. Then
$\boldsymbol{y}$ appears $m$ times in the right-hand side of
\eqref{ineq1a} and $m_{1}$ times in the left-hand side, what proves
the formula \eqref{ineq1a}. If the condition \eqref{ineq1cond1} is
satisfied, then $m_{1} = me^{o(n)}$, from where the equality
\eqref{ineq1b} follows. Note also that by \eqref{defKmy} we have
\begin{equation}\label{ineq1c}
\begin{gathered}
\sum_{i=1}^{M}|D_{\boldsymbol{x}_{i}}(t,p)| =
\sum_{\boldsymbol{y}:m_{p}(\boldsymbol{y}) \geq 1}
\frac{K(\boldsymbol{y},t,p)}{m_{p}(\boldsymbol{y})} =
\sum_{\boldsymbol{y}:m_{p}(\boldsymbol{y}) \geq 1}
m_{t}(\boldsymbol{y}) \sim M2^{h(t)n} -
\sum_{\boldsymbol{y}:m_{p}(\boldsymbol{y}) =0}m_{t}(\boldsymbol{y}).
\end{gathered}
\end{equation}
From the first of the equality \eqref{ineq1c} formulas
\eqref{ineq1a} and \eqref{ineq1b} follow as well. $\qquad \Box$

The formula \eqref{ineq1c} looks simple and attractive, but its
right-hand side has the form ``{\it large minus large}'', what is
not pleasant. Note that in \eqref{ineq1c} we can not neglect the last
sum, because then we get only $r_{\rm crit} \leq 1$, what is not
interesting.

{\bf 2. One more upper bound for $r_{\rm crit}$}. We upperbound
the last sum in в \eqref{ineq1c}  as follows. We have
\begin{equation} \label{Pe7}
\begin{gathered}
\sum\limits_{\boldsymbol{y}:m_{p_{0}}(\boldsymbol{y}) =0}
m_{t}(\boldsymbol{y}) \leq 2^{h(t)n}
|\{\boldsymbol{y}:m_{p_{0}}(\boldsymbol{y}) =0\}|.
\end{gathered}
\end{equation}
Maximum of the cardinality
$|\{\boldsymbol{y}:m_{p_{0}}(\boldsymbol{y}) =0\}|$ is attained when
the code ${\cal C}$ is the ball ${\bf B}_{\boldsymbol{0}}(\tau)$ of
radius $\tau n$, where $r =h(\tau)$. Therefore
\begin{equation} \label{Pe7a}
\begin{gathered}
\max_{{\cal C}}|\{\boldsymbol{y}:m_{p_{0}}(\boldsymbol{y}) =0\}| =
2^{n} - |{\bf B}_{\boldsymbol{0}}(\tau+p_{0})| \sim
2^{h(\tau+ p_{0})n}, \quad \tau+p_{0} \geq 1/2;  \\
\max_{{\cal C}}|\{\boldsymbol{y}:m_{p_{0}}(\boldsymbol{y}) =0\}|
\sim 2^{n}, \quad \tau+p_{0} \leq 1/2.
\end{gathered}
\end{equation}
If $\tau+p_{0} \geq 1/2$, i.e. if $r \geq h(1/2-p_{0})$, then from
\eqref{ineq1c}, \eqref{Pe7} and \eqref{Pe7a} we get
$$
\begin{gathered}
\sum_{i=1}^{M}|D_{\boldsymbol{x}_{i}}(t,p_{0})| \geq
2^{h(t)n}\left[M-2^{h(\tau+p_{0})n}\right] =
2^{h(t)n}\left[2^{h(\tau)n} -2^{h(1-\tau-p_{0})n}\right] \sim
M2^{h(t)n},
\end{gathered}
$$
if $\tau > 1-\tau -p_{0}$, i.e. if $\tau > (1-p_{0})/2$, or,
equivalently, if $r > h[(1-p_{0})/2]$.

Therefore, if
$r \geq \max\{h(1/2-p_{0}), h[(1-p_{0})/2]\} = h[(1-p_{0})/2]$,
then for any $p_{0} \neq p_{1}$ \eqref{lemma1aa} takes the form
$$
\begin{gathered}
F(p_{0},p_{1},r) =
\max_{t> 0}\left\{h(t) +(p_{0}-t)\log\frac{1-p_{1}}{p_{1}}\right\} -
h(p_{0}) = \\
= h(p_{1}) +(p_{0}-p_{1})\log\frac{1-p_{1}}{p_{1}}-h(p_{0})> 0,
\qquad p_{0} \neq p_{1},
\end{gathered}
$$
since maximum over $t$ is attained for $t=p_{1}$.
Therefore, it gives the following upper bound for $r_{\rm crit}$
(weaker than \eqref{Theorem1})
\begin{equation} \label{prop3}
\begin{gathered}
r_{\rm crit}(p_{0},p_{1}) \leq h[(1-p_{0})/2], \qquad
p_{0} \neq p_{1}.
\end{gathered}
\end{equation}

{\it Remark }6. Note that
$1-h(p_{0}) < h(1/2-p_{0}) < h[(1-p_{0})/2]$, $0 < p_{0} < 1/2$.

We improve the bound \eqref{prop3}. In addition to
\eqref{Pe7} we also have
$$
\begin{gathered}
\sum\limits_{\boldsymbol{y}:m_{p_{0}}(\boldsymbol{y}) =0}
m_{t}(\boldsymbol{y}) \leq
M|\{\boldsymbol{y}:m_{p_{0}}(\boldsymbol{y}) =0\}|.
\end{gathered}
$$
Therefore, if $\tau+p_{0}\geq 1/2$ and $t\geq 1-\tau -p_{0}$, then
$$
\begin{gathered}
\sum_{i=1}^{M}|D_{\boldsymbol{x}_{i}}(t,p_{0})| \geq
M\left[2^{h(t)n}-2^{h(1-\tau-p_{0})n}\right] \sim M2^{h(t)n}.
\end{gathered}
$$
By \eqref{dist1g}-\eqref{dist1g1} it is necessary to have
\begin{equation} \label{Pe71a}
\begin{gathered}
\max_{t \geq 1-\tau -p_{0}}f(t,p_{0},p_{1}) > 0, \\
f(t,p_{0},p_{1})= h(t)+(p_{0}-t)\log\frac{1-p_{1}}{p_{1}}- h(p_{0}).
\end{gathered}
\end{equation}
Maximum of the function $f(t,p_{0},p_{1})$ over
$t \geq 1-\tau -p_{0}$ is attained for
$t=\max\{p_{1},1-\tau -p_{0}\}$, since
\begin{equation} \label{Pe71a1}
\begin{gathered}
\max_{t}f(t,p_{0},p_{1}) = f(p_{1},p_{0},p_{1}) > 0, \quad
p_{0} \neq p_{1};  \qquad f(p_{0},p_{0},p_{1}) = 0, \\
f'_{t}(t,p_{0},p_{1})= \log\frac{1-t}{t}- \log\frac{1-p_{1}}{p_{1}},
\qquad f''_{tt}(t,p_{0},p_{1}) < 0, \\
\rm{sign } \ f'_{t}(t,p_{0},p_{1}) = \rm{sign} \ (p_{1}-t).
\end{gathered}
\end{equation}

1) Therefore, if $p_{1} \geq 1-\tau -p_{0}$, then from
\eqref{Pe71a}-\eqref{Pe71a1} for $p_{0} \neq p_{1}$ we get
\begin{equation} \label{Pe72a}
\begin{gathered}
\max_{t \geq 1-\tau -p_{0}}f(t,p_{0},p_{1}) = h(p_{1}) +
(p_{0}-p_{1})\log\frac{1-p_{1}}{p_{1}}-h(p_{0}) > 0.
\end{gathered}
\end{equation}
Hence if
$\tau \geq \max\{1/2-p_{0},1-p_{0} -p_{1}\} = 1-p_{0} -p_{1}$,
then for $p_{0} \neq p_{1}$ the inequality \eqref{Pe72a} holds,
from where the estimate follows
\begin{equation} \label{Pe72b}
\begin{gathered}
\tau_{\rm crit} \leq 1-p_{0} -p_{1}, \qquad
r_{\rm crit}=h(\tau_{\rm crit}).
\end{gathered}
\end{equation}

2) If $p_{1} < 1-\tau -p_{0}$, then maximum in \eqref{Pe71a} is
attained for $t=1-\tau -p_{0}$, and then
$$
\begin{gathered}
\max_{t \geq 1-\tau -p_{0}}f(t,p_{0},p_{1}) =
f(1-\tau -p_{0},p_{0},p_{1}).
\end{gathered}
$$
Note that
$$
\begin{gathered}
f(p_{0},p_{0},p_{1}) = 0, \qquad f'_{t=p_{0}}(t,p_{0},p_{1}) \neq 0,
\quad p_{0} \neq p_{1}; \\
f''_{tt}(t,p_{0},p_{1}) < 0, \qquad
\rm{sign } \ f'_{t}(t,p_{0},p_{1}) = \rm{sign} \ (p_{1}-t).
\end{gathered}
$$
Let also $p_{0} > 1-\tau -p_{0}$ (i.e. $\tau > 1-2p_{0}$). Then
$\max\limits_{t \geq 1-\tau -p_{0}}f(t,p_{0},p_{1}) > 0$
(it is sufficient to set $t$, close to $p_{0}$). Therefore
\begin{equation} \label{Pe72ca}
\begin{gathered}
\tau_{\rm crit} \leq 1-2p_{0}, \qquad
r_{\rm crit}=h(\tau_{\rm crit}).
\end{gathered}
\end{equation}

As a result, from \eqref{Pe72b} and \eqref{Pe72ca} we get

{P r o p o s i t i o \,3}.
{\it For any $p_{0},p_{1} \in [0,1/2]$ for $r_{\rm crit}$ the
upper bound holds }
\begin{equation} \label{prop3a}
\begin{gathered}
\tau_{\rm crit}(p_{0},p_{1})
\leq \min\left\{1-p_{0} -p_{1}, 1-2p_{0}\right\}, \qquad
r_{\rm crit}=h(\tau_{\rm crit}).
\end{gathered}
\end{equation}

C o r o l l a r y. {\it If $p_{0}=1/2$, then from \eqref{prop3a}
it follows $\tau_{\rm crit}(1/2,p_{1})  r_{\rm crit}(1/2,p_{1})= 0$}.

Earlier that particular result was proved by different method in
\cite[предложение 3]{BAH1}. Also the best exponent
$e_{\rm d}(\gamma, r)$ for $\gamma \geq 0$, $0 \leq r \leq 1$ from
\eqref{def2a} was obtained there.

\begin{center}
{\large\bf \S\;6. ``Potential'' additive upper bound for
$r_{\rm crit}$}.
\end{center}

Theorem 1 was proved replacing in the formula \eqref{lemma1} the
exponential number $M$ of codewords $\{\boldsymbol{x}_{i}\}$ by
two closest codewords $(\boldsymbol{x}_{i},\boldsymbol{x}_{j})$.
Such method gives optimal results only if it is possible to choose
a pair $(\boldsymbol{x}_{i},\boldsymbol{x}_{j})$ with
$d(\boldsymbol{x}_{i},\boldsymbol{x}_{j}) = \omega n$ and small
$\omega > 0$. In the problem statement considered we can not do
that.

In order to strengthen Theorem 1 it is necessary to consider in
\eqref{lemma1} an exponential number $M$ of codewords
$\{\boldsymbol{x}_{i}\}$, what is much more difficult
\cite{Bur15,Bur6,Bur5}. We strengthen Theorem 1 provided it is
possible to use in the formula \eqref{lemma1} an additive
approximation.

We assume that for all $\{\boldsymbol{x}_{i}\}$ in the formula
\eqref{lemma1} the additive approximation holds as $n \to \infty$
\begin{equation} \label{approx1}
\begin{gathered}
{\mathbf P}
\left\{\bigcup_{k \neq i}SL_{\boldsymbol{x}_k}(p_{0},\delta)\Big|
p_{1},\boldsymbol{x}_{i}\right\} = 2^{o(n)}\sum_{k \neq i}
{\mathbf P}\left\{SL_{\boldsymbol{x}_k}(p_{0},\delta)\Big|
p_{1},\boldsymbol{x}_{i}\right\}.
\end{gathered}
\end{equation}

Then (see \eqref{geom2aa}) with
$d(\boldsymbol{x}_i,\boldsymbol{x}_k) = \omega_{ik}n$
$$
\begin{gathered}
{\mathbf P}\left\{\bigcup_{k \neq i}
SL_{\boldsymbol{x}_k}(p_{0},\delta)\Big|
p_{1},\boldsymbol{x}_{i}\right\} = 2^{o(n)}
\sum_{k \neq i}2^{f(p_{0},p_{1},\omega_{ik})n}
\end{gathered}
$$
and
\begin{equation} \label{lemma1ab}
\begin{gathered}
\sum_{i=1}^{M}{\mathbf P}
\left\{\bigcup_{k \neq i}SL_{\boldsymbol{x}_k}(p_{0},\delta)\Big|
p_{1},\boldsymbol{x}_{i}\right\} = 2^{o(n)}\sum_{i=1}^{M}
\sum_{k \neq i}2^{f(p_{0},p_{1},\omega_{ik})n}.
\end{gathered}
\end{equation}

In order to develop relations \eqref{lemma1ab}, introduce some
additional notions.

{\it Code spectrum} ({\it distance distribution}) of length $n$ code
${\cal C}$ is the $(n+1)$--tuple
$B({\cal C}) = (B_{0},B_{1},\ldots,B_{n})$ with components
\begin{equation}\label{defBi}
B_{i} = \left|{\cal C}\right|^{-1}\left|\left\{
(\boldsymbol{x},\boldsymbol{y}): \boldsymbol{x},\boldsymbol{y}
\in {\cal C},\,d(\boldsymbol x,\boldsymbol{y}) = i\right\}\right|,
\qquad i=0,1,\ldots,n.
\end{equation}

In other words, $B_{i}$ is average number of codewords
$\boldsymbol{y}$ on the distance $i$ from the codeword
$\boldsymbol x$. The total number of ordered codepairs
$\boldsymbol{x},\boldsymbol{y} \in {\cal C}$ with
$d(\boldsymbol x,\boldsymbol{y}) = i$ equals $|{\cal C}|B_{i}$.
Denote also $B_{\omega n} = 2^{b(\omega,r)n}$.

Then we can continue the formula \eqref{lemma1ab} as follows
$$
\begin{gathered}
\sum_{i=1}^{M}{\mathbf P}
\left\{\bigcup_{k \neq i}SL_{\boldsymbol{x}_k}(p_{0},\delta)\Big|
p_{1},\boldsymbol{x}_{i}\right\}
= 2^{o(n)}M\sum_{\omega>0}2^{[b(\omega,r)+f(p_{0},p_{1},\omega)]n}.
\end{gathered}
$$
Therefore (see \eqref{geom2aa}-\eqref{geom2aa1})
\begin{equation} \label{geom2aa4}
\begin{gathered}
\frac{1}{n}\log\left[\sum_{i=1}^{M}{\mathbf P}
\left\{\bigcup_{k \neq i}SL_{\boldsymbol{x}_k}(p_{0},\delta)\Big|
p_{1},\boldsymbol{x}_{i}\right\}\right] = r + \max_{\omega,t}
\left\{b(\omega,r)+ f(p_{0},p_{1},\omega,t)\right\} + o(1),
\end{gathered}
\end{equation}
where $f(p_{0},p_{1},\omega,t)$ is defined in \eqref{geom2aa1}.
Then for the function $F(p_{0},p_{1},r)$ from \eqref{lemma1aa}
and \eqref{geom2aa4} we have
\begin{equation} \label{lemma2a}
\begin{gathered}
F(p_{0},p_{1},r)  = \max_{\omega,t}\left\{b(\omega,r) +
g(p_{0},t,\omega) + (p_{0}-t)\log \frac{1-p_{1}}{p_{1}}-h(p_{0})
\right\}.
\end{gathered}
\end{equation}

As an estimate for $b(\omega,r)$ in \eqref{lemma2a} we use
a function $b_{\rm low}(\omega,r)$ with the following property:
there exists a value $\omega_{\rm max}=\omega_{\rm max}(r) > 0$,
such that
\begin{equation} \label{blow1}
\begin{gathered}
\max_{0 < \omega \leq \omega_{\rm max}}
\left[b(\omega,r) -b_{\rm low}(\omega,r)\right] \geq 0, \qquad r>0.
\end{gathered}
\end{equation}

Then in order the inequality $F(p_{0},p_{1},r) > 0$
(see \eqref{lemma1a}) be valid, it is sufficient the following
condition (see \eqref{geom2aa1} and \eqref{lemma2a}) be satisfied
\begin{equation} \label{equalPe1a1}
\begin{gathered}
\min_{0 < \omega \leq \omega_{\rm max}}\max_{t> 0}
\left\{b_{\rm low}(\omega,r) + g(p_{0},t,\omega) +
(p_{0}-t)\log \frac{1-p_{1}}{p_{1}}-h(p_{0})\right\} > 0.
\end{gathered}
\end{equation}

We use in \eqref{equalPe1a1} as $b_{\rm low}(\omega,r)$ the best of
known such functions $\mu(r,\alpha,\omega)$,
$h_{2}(\tau) = h_{2}(\alpha)-1+r$, with arbitrary
$\alpha \in [\delta_{GV}(r),1/2]$ (see \eqref{defGV}, \eqref{spectr1}
and Theorem 2 in Appendix). The function $\mu(r,\alpha,\omega)$
satisfies the condition \eqref{blow1}. Moreover, it monotonically
increases in $r$ and $\omega_{\rm max} = G(\alpha,\tau)$, where
$G(\alpha,\tau)$ is defined in \eqref{defG}. Then in order the
inequality \eqref{equalPe1a1} be satisfied, it is sufficient the
condition be fulfilled
\begin{equation} \label{equalPe1a9}
\begin{gathered}
\min_{0 < \omega \leq \omega_{\rm max}}\max_{t> 0}
K(p_{0},p_{1},r,\omega,t) > 0,
\end{gathered}
\end{equation}
where
\begin{equation} \label{DefG4}
\begin{gathered}
K(p_{0},p_{1},r,\omega,t) = \mu(r,p_{0},\omega) +
g(p_{0},t,\omega) + (p_{0}-t)\log\frac{1-p_{1}}{p_{1}}-h(p_{0}).
\end{gathered}
\end{equation}
Note that $K(p_{0},p_{1},r,0,p_{0})=0$. In order to avoid bulky
calculations, we set $t=p_{0}$. The function
$K(p_{0},p_{1},r,\omega,p_{0})=0$ is concave in $\omega$, i.e.
$K''(p_{0},p_{1},r,\omega,p_{0})_{\omega\omega} < 0$ (the simplest
way is to check that with Maple). Therefore, minimum over $\omega$
is attained for $\omega =\omega_{\rm max} =G(\alpha,\tau)$ and
it is sufficient to check the condition \eqref{equalPe1a9} for
$\omega  = G(\alpha,\tau)$. The following useful formula
\cite[Lemma 4]{Bur15} is known:
\begin{equation}\label{ident1ab}
\begin{gathered}
\mu(r,\alpha,G(\alpha,\tau)) = h_{2}(G(\alpha,\tau)) + r -1, \qquad
h_{2}(\alpha) - h_{2}(\tau)=1-r.
\end{gathered}
\end{equation}

Consider only more simple

{\bf Case $p_{1} < p_{0} \leq 1/2$}. Set
$r = r_{0}=1-h(p_{0})$ and $\alpha = p_{0}$
(then $\delta_{GV}(r_{0}) = p_{0}$, $\tau = 0$). We have
$G(\alpha,\tau) = 2p_{0}(1-p_{0})$ and it is sufficient to check
the condition \eqref{equalPe1a9} for $\omega  = 2p_{0}(1-p_{0})$.
From \eqref{DefG4}-\eqref{ident1ab} with $\alpha = p_{0}$, $\tau =0$,
$r=r_{0}=1-h(p_{0})$, $t=p_{0}$ and
$\omega_{\rm max} =G(\alpha,\tau) = 2p_{0}(1-p_{0})$ we have
$$
\begin{gathered}
K(p_{0},p_{1},1-h(p_{0}),\omega_{\rm max},p_{0}) =
h_{2}(\omega_{\rm max})+ g(p_{0},p_{0},\omega_{\rm max})-2h(p_{0}),
\end{gathered}
$$
where
$$
\begin{gathered}
g(p,p,2p(1-p)) = 2p(1-p) +
[1-2p(1-p)]h\left[\frac{p^{2}}{1-2p(1-p)}\right].
\end{gathered}
$$
It is possible to check that for $\omega_{0} = 2p_{0}(1-p_{0})$
the equality holds
\begin{equation} \label{equalPe1a3}
\begin{gathered}
K(p_{0},p_{1},1-h(p_{0}),\omega_{0},p_{0})  = h_{2}(\omega_{0}) +
\omega_{0} + (1-\omega_{0})
h\left(\frac{p_{0}^{2}}{1-\omega_{0}}\right)-2h(p_{0}) = 0.
\end{gathered}
\end{equation}
We also have
\begin{equation} \label{equalPe1a4}
\begin{gathered}
\left[K(p_{0},p_{1},1-h(p_{0}),\omega_{0},t)\right]'_{t} =
\frac{1}{2}\log\frac{(1-t)^{2}-(1-\omega_{0}-p_{0})^{2}}
{t^{2}-(\omega_{0}-p_{0})^{2}} -\log\frac{1-p_{1}}{p_{1}}, \\
\left[K(p_{0},p_{1},1-h(p_{0}),\omega_{0},t)\right]''_{tt} < 0.
\end{gathered}
\end{equation}
Therefore, for $t=p_{0}$ we have
\begin{equation} \label{equalPe1a5}
\begin{gathered}
\left[K(p_{0},p_{1},1-h(p_{0}),\omega_{0},t)\right]'_{t=p_{0}} =
\log\frac{1-p_{0}}{p_{0}} -\log\frac{1-p_{1}}{p_{1}} < 0, \qquad
p_{1} < p_{0},
\end{gathered}
\end{equation}
It follows from \eqref{equalPe1a3}-\eqref{equalPe1a5} that
$$
\begin{gathered}
K(p_{0},p_{1},1-h(p_{0}),\omega_{0},t) > 0, \qquad t < p_{0}.
\end{gathered}
$$
Therefore, the inequality \eqref{equalPe1a9} holds for any
$r > r_{0}=1-h(p_{0})$ и $p_{1} < p_{0} \leq 1/2$.

As a result, we get the conditional result:

{P r o p o s i t i o n \,4}.
{\it If the additive approximation \eqref{approx1} holds, then
$r_{\rm crit}(p_{0},p_{1}) = 1-h(p_{0})$,
$0 < p_{1} < p_{0} \leq 1/2$.}

{\it Remark }6. It is possible to show that Theorem 1 and the
formula \eqref{Theorem1} hold for any $p_{1} < p_{0} \leq 1/2$.
For that purpose we can perform similarly to \cite{Bur15}, using
Lemma 2 and considering separately the case of equality in the
formula \eqref{ineq1a} (essentially, it is equivalent to the
considered in \S 6 case), and the case of inequality in the
formula \eqref{ineq1a}. Proof in the second case turns out to be
too bulky (and oriented only to the binary channel {\rm BSC}$(p)$).
For that reason we omit that proof. Certainly, there should be
a simpler proof.

\hfill {\large\sl APPENDIX}

{\bf 1}. F u n c t i o n \ $g(t,p,\omega)$ \ a n d \ f o r m u l a \
\eqref{defD3a}. Consider codewords
$\boldsymbol{x}=\boldsymbol{0}$ and $\boldsymbol{x}_{1}$ with
$d(\boldsymbol{x},\boldsymbol{x}_{1}) =
w(\boldsymbol{x}_{1}) = \omega n$, and the set
$S_{\boldsymbol{x},\boldsymbol{x}_{1}}(t,p,\omega)$ from
\eqref{defD4}. We may assume that
$\boldsymbol{x}_{1}=(1,\ldots,1,0,\ldots,0)$ and
has first $\omega n$ ``ones'', and then $(1-\omega)n$ ``zeros''.
Let also
$\boldsymbol{u}\in S_{\boldsymbol{x},\boldsymbol{x}_{1}}(t,p,\omega)$
has $u_{1}n$ ``ones'' on the first $\omega n$ positions, and then
$u_{2}n$ ``ones'' on the next $(1-\omega)n$ positions. Since
$u_{1}+u_{2}=t$, $\omega-u_{1}+u_{2}=p$, then
\begin{equation} \label{geom2a1}
\begin{gathered}
u_{1} = \frac{t-p+\omega}{2}, \qquad u_{2} = \frac{t+p-\omega}{2},
\end{gathered}
\end{equation}
and as $n \to \infty$ we get
\begin{equation} \label{geom2a}
\begin{gathered}
\frac{1}{n}\log|S_{\boldsymbol{x},\boldsymbol{x}_{1}}(t,p,\omega)| =
\frac{1}{n}\log\left[
\binom{\omega n}{u_{1}n}\binom{(1-\omega)n}{u_{2}n}\right] = \\
= \omega h\left(\frac{u_{1}}{\omega}\right) +
(1-\omega)h\left(\frac{u_{2}}{1-\omega}\right) + o(1) =
g(t,p,\omega)+ o(1),
\end{gathered}
\end{equation}
where
\begin{equation} \label{defg}
\begin{gathered}
g(t,p,\omega) = \omega h\left(\frac{t+\omega-p}{2\omega}\right) +
(1-\omega)h\left(\frac{t+p-\omega}{2(1-\omega)}\right).
\end{gathered}
\end{equation}

We also have
\begin{equation} \label{geom2b}
\begin{gathered}
2g'_{\omega}(p,t,\omega) =
-2\log\frac{1-\omega}{\omega} +
\log\frac{(1-\omega)^{2}-(1-t-p)^{2}}{\omega^{2}-(t-p)^{2}}, \\
2g'_{t}(p,t,\omega) =
\log\frac{(1-t)^{2}-(1-\omega-p)^{2}}{t^{2}-(\omega-p)^{2}}, \qquad
g''_{tt}(p,t,\omega) < 0, \qquad
g''_{\omega\omega}(p,t,\omega) \leq 0.
\end{gathered}
\end{equation}
For the root $\omega_{0}$ of the equation
$g'_{\omega}(t,p,\omega) =0$ we have
\begin{equation} \label{geom2d}
\begin{gathered}
\omega_{0} = \frac{p-t}{1-2t}, \qquad g(t,p,\omega_{0}) = h(t).
\end{gathered}
\end{equation}

{\bf 2}. F u n c t i o n \ $\mu(R,\alpha,\omega)$. Introduce the
function \cite{M4} ($0\leq \tau\leq \alpha \leq 1/2$)
\begin{equation}\label{defG}
G(\alpha,\tau) = 2\frac{\alpha(1-\alpha) - \tau(1-\tau)}
{1+2\sqrt{\tau(1-\tau)}} \geq 0.
\end{equation}
For $\alpha,\tau$, such that $0 \leq \tau \leq \alpha \leq 1/2$ and
$h_{2}(\alpha) - h_{2}(\tau)=1-R$, introduce the function
\cite{L1}
\begin{equation}\label{defmu1}
\begin{gathered}
\mu(R,\alpha,\omega) = h_{2}(\alpha)- 2\int\limits_{0}^{\omega/2}
\log \frac{P + \sqrt{P^{2}-4Qy^{2}}}{Q}\,dy - (1-\omega)
h_{2}\left(\frac{\alpha - \omega/2}{1-\omega}\right), \\
P = \alpha(1-\alpha) - \tau(1-\tau)-y(1-2y), \qquad
Q = (\alpha - y)(1-\alpha-y).
\end{gathered}
\end{equation}

Denote the function $\delta_{GV}(R) \leq 1/2$
(Varshamov - Gilbert bound) as
\begin{equation}\label{defGV}
1 - R = h_{2}(\delta_{GV}(R)), \qquad 0 \leq R \leq 1.
\end{equation}

Importance of the function $\mu (R,\alpha,\omega)$ and its relation
to the code spectrum $\{B_{i}\}$ (see \eqref{defBi}) is described
by the following variant of Theorem 3 from \cite{Bur19}.

T h e o r e m \,2 \cite[Theorem 3]{Bur19}.
{\it For any $(R,n)$-code and any $\alpha \in [\delta_{GV}(R),1/2]$
there exist $r_{1}(R,\alpha) > 0$ and $\omega$,
 $0 < r_{1}(R,\alpha) \leq \omega \leq G(\alpha,\tau)$,
where $h_{2}(\tau) = h_{2}(\alpha)-1+R$, and $G(\alpha,\tau)$ is
defined in \eqref{defG}, such that
\begin{equation}\label{spectr1}
n^{-1}\log B_{\omega n} \geq \mu(R,\alpha,\omega) + o(1),
\qquad n \to \infty.
\end{equation}
For $\mu(R,\alpha,\omega)$ from \eqref{defmu1} the non-integral
representation \eqref{reprmu1}-\eqref{auxdefmu01} also holds}.

{\it Remark }7. Theorem 2 makes more precise Theorem 5 from
\cite{L1} (see also \cite[Theorem 2]{Bur6}. With $r_{1}=0$
Theorem 2 turns into Theorem 5 from \cite{L1}.
In \cite[теорема 3]{Bur19} there are estimates for
$r_{1}(R,\alpha) > 0$.

P r o p o s i t i o n \,5 \cite[Proposition 3]{Bur15}.
{\it For the function $\mu(R,\alpha,\omega)$ the representation
holds
\begin{equation}\label{reprmu1}
\begin{gathered}
\mu(R,\alpha,\omega) =(1-\omega)h_{2}\left(\frac{\alpha- \omega/2}
{1-\omega}\right) -h_{2}(\alpha) + 2h_{2}(\omega) +
\omega\log\frac{2\omega}{e} - T(A,B,\omega),
\end{gathered}
\end{equation}
where
\begin{equation}\label{auxdefmu0}
\begin{gathered}
T(A,B,\omega) = \omega\log(v-1)-
(1-\omega)\log\frac{v^{2} - A^{2}}{v^{2} - B^{2}} + \\
+ B\log\frac{v+B}{v-B}- A\log\frac{v+A}{v-A} -
\frac{(v-1)(B^{2} - A^{2})}{(v^{2} - B^{2})\ln 2}, \\
v = \frac{\sqrt{B^{2}\omega^{2} - 2a_{1}\omega +
a_{1}^{2}} + a_{1}}{\omega}, \qquad a_{1} = \frac{B^{2}-A^{2}}{2}.
\end{gathered}
\end{equation}
and }
\begin{equation}\label{auxdefmu01}
\begin{gathered}
h_{2}(\alpha)-h_{2}(\tau) = 1 - R, \quad A = 1-2\alpha, \quad
B = 1 -2\tau, \quad 0 \leq \tau \leq \alpha \leq 1/2.
\end{gathered}
\end{equation}

We have for any $\alpha_{0}(R) \leq \alpha < 1/2$ and
$\omega > 0$
$$
\frac{d\mu(R,\alpha,\omega)}{d\alpha} > 0, \qquad
\alpha_{0}(R)= h_{2}^{-1}(1-R).
$$
For any $\alpha > 0$ and $R > 0$ we also have $\mu(R,\alpha,0) = 0$
and $\mu'_{\omega}(R,\alpha,\omega)\Big|_{\omega = 0} > 0$.
Moreover, for any $0 \leq \tau \leq \alpha \leq 1/2$ and
$0 < \omega < G(\alpha,\tau)$
$$
\mu''_{\omega^{2}}(R,\alpha,\omega) > 0.
$$
3) For any $\omega > 0$ we have  $\mu(0,1/2,\omega) = 0$.

\begin{center}{ACKNOWLEDGEMENTS}\end{center}
The author appreciates Shun Watanabe and the reviewer for useful
discussions and constructive critical remarks which improved the paper.


\end{document}